# Modeling of hydrothermal aging of short flax fiber reinforced composites

Arnaud Regazzi, Romain Léger *, Stéphane Corn, Patrick Ienny

*École des Mines d'Alès, C2MA, 6 avenue de Clavières, F-30319 ALES Cedex, France*

**a b s t r a c t**

As a contribution to the prediction of the evolutionary behavior of biocomposites in service conditions, this study focused on the simulation of the hydrothermal aging of short natural fiber reinforced composites made by extrusion/injection molding. We endeavored to model the reversible modifications of the behavior of PLA and PLA/flax composites when immersed in water at different temperatures (20, 35 and 50 °C). A numerical model accounting for the heterogeneous mechanisms involved during aging such as water diffusion and the resulting swelling and plasticizing of polymers was implemented. Simulated data proved to be in perfect accordance with experimental results as long as no irreversible mechanism was occurring. The deviations of the simulated data from experimental results were limited at 35 °C but significant at 50 °C. Finally, the influence of moisture on the local elastic modulus of flax fibers was inferred thanks to the Halpin-Kardos homogenization model.

## 1. Introduction

### 1.1. Context

The demand for bio-based materials in semi-structural and structural application is constantly growing to conform to new environmental policies enacted in Europe and worldwide which try to replace conventional oil-based polymers and composites. Natural fibers reinforced composites and particularly fax fibers reinforced composites meet an important success because of their interesting specific mechanical properties, availability and reasonable price compared to conventional glass fiber reinforced composites [1]. When employed with a bio-based matrix such as PLA, these fibers form a biocomposite which is fully recyclable. However, the hydrophilic nature of these materials results in a poor durability when exposed to humidity and temperature [2–5] restraining the adoption of these biocomposites. In order to improve the behavior of biocomposites in humid environment, much work has been dedicated to the chemical modification of natural fiber [6–8]. Besides, the evolution of their mechanical properties in service conditions can be predicted with numerical simulations that rely on experimental results [9,10]. Having a numerical model offers the opportunity to assess the influence of many parameters during aging, including the temperature, the fiber ratio or the structure geometry. The purpose of this paper is to present the development and the results of a numerical model that enables to predict the long term properties of biocomposites with various fiber contents and different hydrothermal aging conditions.

### 1.2. Methodology

The complexity of the various physico-chemical processes encountered and analyzed on PLA/flax composites prevents any simple analytical modeling. In order to propose a solution as a part of a more complete model, it was decided to focus this paper on modeling the reversible phenomena (*i.e.* those which do not induce a permanent modification of the properties of the material after desiccation) caused by a thermo-hydric aging [11]. As a result, only swelling (*i.e.* the dimensional changes caused by the absorption of a solvent) and plasticizing (*i.e.* the mechanical softening caused by the absorption of a solvent) were taken into account. Hence, irreversible processes, like hydrolysis, chains relaxation, crystallization, cavitation, crazing, and fiber/matrix debonding, were excluded from the proposed model. It was chosen to consider the evolution of the local elastic properties of this material with the water concentration because of the high sensitivity of its stiffness to moisture [11].

In order to take into account the three dimensional geometry of dog-bone shaped samples that were used to measure ultimate mechanical properties after aging, a numerical model was developed. Indeed, the moisture diffusion that occurs inside the sample is conditioned by simultaneous fluxes across the different surfaces exposed to water. They induce complex water concentration gradients inside the volume, prohibiting the use of a simple analytical approach. Despite the prominent areas of inferior and superior surfaces, considering a unidirectional sorption would thus induce a significant error in the parameters identification [12]. The finite element analysis software Comsol Multiphysics® was used to develop a thermo-hydro-mechanical constitutive model thanks to its ability to solve coupled physics problems. The coefficients of this model were assessed on the basis of experimental measurements at the macroscopic scale of the studied structure.

* Corresponding author.
  *E-mail address:* romain.leger@mines-ales.fr (R. Léger).

As a result, the model was intended to simulate the time evolution of the physical and mechanical behavior of PLA/flax composites during a non-damaging (reversible) thermo-hydric aging.

## 2. Materials and methods

### 2.1. Materials

#### 2.1.1. Poly(lactic acid) granules

PLA Ingeo™ 7000D resin was obtained from NatureWorks® LLC (Blair, NE, USA). These granules were designed for injection stretch blow molded applications. According to the producer, this grade of PLA had a density of 1.24 g/cm³, a glass transition temperature between 55 and 60 °C and a melting temperature between 155 and 165 °C.

#### 2.1.2. Flax fibers

The short flax fibers (*Linum usitatissimum*) FIBRA-S® 6A used for this study were provided by Fibres Recherches Développement® (Troyes, France). According to the technical data sheet, fibers (bundles) were 6 mm long with a diameter of 260 ± 150 μm and their density was between 1.4 and 1.5 g/cm³. Concerning quasi-static mechanical properties, Young's modulus of bundles was 36 ± 13 GPa, maximum stress was 750 ± 490 MPa and strain at break was 3.0 ± 1.9%.

### 2.2. Experimental techniques

#### 2.2.1. Processing

Various fiber weight contents were used: 0% (neat PLA) hereafter named PLA, 10% hereafter named PLA-F10, and 30% hereafter named PLA-F30. Polylactic acid granules and flax fibers were dried at 80 °C for at least 24 h and under vacuum at 120 °C for 4 h respectively. Composite granules were obtained with a corotative twin-screw extruder (Clextral BC21, screw length = 900 mm; temperature profile along the screw and at the die = 180 °C). After a second drying step under vacuum at 80 °C during 24 h, compounded granules were molded with an injection molding machine (Krauss Maffei KM50-180CX) into dog-bone samples whose geometry meets the requirements of the standard ISO 527-2 1BA (Fig. 1). The temperature profile was increasing up to 200 °C and the mold was kept at 25 °C. After processing, samples were stored at room temperature and 2%rh (relative humidity) before characterization or aging. This equilibrium state constitutes the reference for evaluating the effects of aging on these materials.

#### 2.2.2. Morphology

Fiber size and shape distributions were measured following the methodology described by Le Moigne et al. [13]. A small amount of composites (0.5 g) was dissolved in 20 mL of chloroform during 30 min at ambient temperature and under gentle mixing using a magnetic stirrer. After complete dissolution, a droplet was deposited between a glass slide and a glass lamella. Fibers were observed with a Leica Laborlux 11 POLS optical microscope in transmission mode equipped with a 3-CCD camera (JVC KY-F55BE). Crossed polarizers helped enhancing the contrast between fibers and background. The pictures obtained were analyzed with the software Aphelion® V4.3.1 (ADCIS and Amerinex Applied Imaging) which detects the contour of fibers and evaluates the size and shape distributions. About 2000 single fibers were considered from 100 images. Crossing fibers were discarded from this analysis as well as particles with length smaller than 10 μm because of the pictures resolution (1.72 μm/pixel).

Considering that the Feret diameter of an object is defined as the distance between two parallel tangential lines, for a given fiber, the maximal Feret diameter was used as its length, while the minimal Feret diameter was used as its width. Its shape factor $\phi$ was defined as the ratio between maximal Feret diameter and minimal Feret diameter. Distributions of length and shape factor are displayed in number for each material in Fig. 2. It was observed that fibers from PLA-F10 were longer than from PLA-F30. Actually the length of 75% of the fibers in PLA-F10 ranged between 20 and 70 μm with a median of 45 μm, while in PLA-F30 their length ranged between 18 and 50 μm with a median of 30 μm. The shape factor of half of the fibers in PLA-F10 ranged between 2 and 4 with a median close to 3, while in PLA-F30, their shape factor ranged between 1.8 and 3.4 with a median of 2.2. These results are explained by the higher shear stresses occurring in the extruder with increasing the fiber content in the compound, which leads to shorter fibers.

#### 2.2.3. Aging

Three series of isothermal water immersion experiments were conducted for each material (PLA, PLA-F10 and PLA-F30) at 20, 35, and 50 °C. For each temperature and material, 10 samples were periodically removed from water, slightly wiped, characterized at room temperature (*ex situ* approach) and then re-immersed, up to 51 days. These characterization tests consisted in measuring the mass, the dimensions, and the dynamic elastic modulus of samples. The dynamic elastic modulus, which corresponds to the conservative modulus evaluated from forced vibrations, was assessed from the first natural flexural vibration mode and by a mean of a numerical model. Details about all experimental methods and their results can be found in [11].

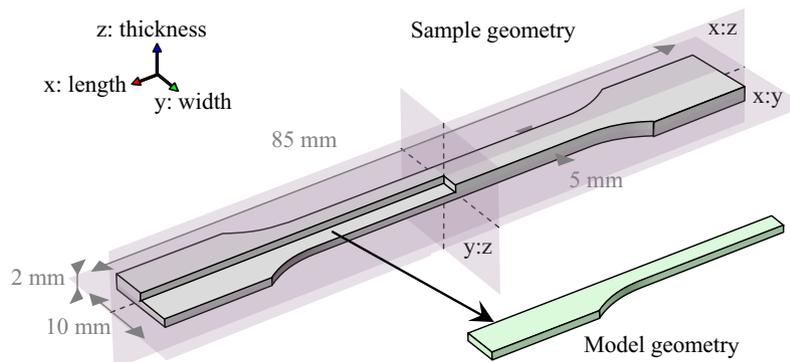

**Fig. 1.** Geometry of samples and the geometry used for the model implementation.

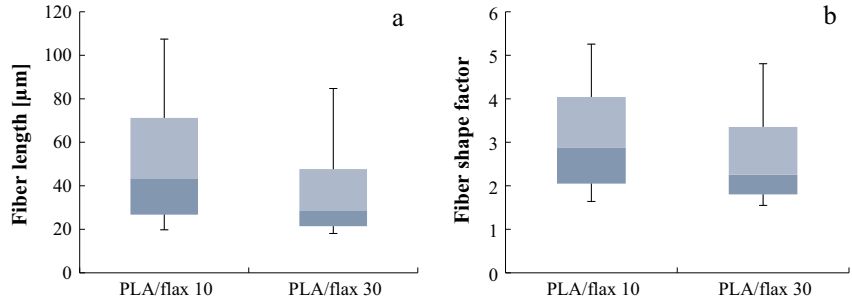

**Fig. 2.** Distribution of fiber length (a) and fiber shape factor (b) in number for composites (boxplots indicate upper decile, upper quartile, median, lower quartile and lower decile).

**Table 1**
Homogeneous water contents obtained at saturation for a variety of temperature/humidity combinations.

| Temperature (°C) | Environment | Water content at saturation [%] | | |
|---|---|---|---|---|
| | | PLA | PLA-F10 | PLA-F30 |
| 20 | 97%rh (K$_2$SO$_4$) | 0.81 | 2.01 | 4.50 |
| 35 | 50%rh (K$_2$CO$_3$) | 0.32 | 0.91 | 1.16 |
| 35 | 97%rh (K$_2$SO$_4$) | 0.82 | 2.23 | 5.26 |
| 20 | Water | 0.93 | 1.86 | 6.00 |
| 35 | Water | 0.93 | 2.89 | 8.44 |
| 50 | Water | 1.20 | 5.08 | 12.4 |

Among the results, each immersion aging test provided a dynamic elastic modulus for a homogeneous concentration (*i.e.* at saturation when water uptake stops increasing). The water contents reached at saturation for each material and each temperature are gathered in Table 1.

Besides, in order to provide additional experimental data to the numerical model, other samples were exposed to different controlled atmospheric conditions, since the water content at saturation depends on the environmental conditions. The water concentration in wet air being lower than in immersion, these conditions allowed achieving lower water contents at saturation. To perform such aging, air relative humidity (%rh) was regulated thanks to potassium carbonate (K$_2$CO$_3$) and potassium sulfate (K$_2$SO$_4$) salts laid in desiccators. The temperature was controlled by placing desiccators in a temperature-controlled room at 20 °C and in oven at 35 °C. The different temperature/humidity combinations applied and the subsequent water content at saturation are summarized in Table 1. Similarly to immersion tests, the dynamic elastic modulus was evaluated at room temperature from free vibrations tests [11]. The dynamic elastic moduli of samples subjected either to immersion or wet air were plotted as a function of their water content in Fig. 3.

## 3. Models

For simplification purpose, the proposed model assumed materials to be homogeneous and temperature was considered to have an impact only on the diffusion process. As the specifications chosen for the model were limited to the reversible processes, hydric swelling and plasticizing were the only phenomena taken into account. Of course, they were both completely dependent of the water diffusion kinetic into the material.

### 3.1. Water sorption/diffusion

Sorption experimental results showed three kinds of kinetics (cf. Fig. 5):

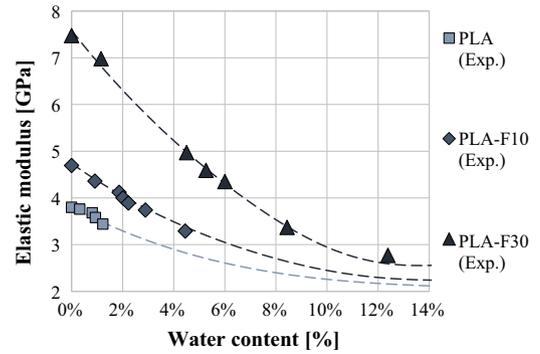

**Fig. 3.** Experimental values (markers) and interpolated function (dashed lines) of the local dynamic elastic modulus of PLA and flax/PLA composites as a function of the water content.

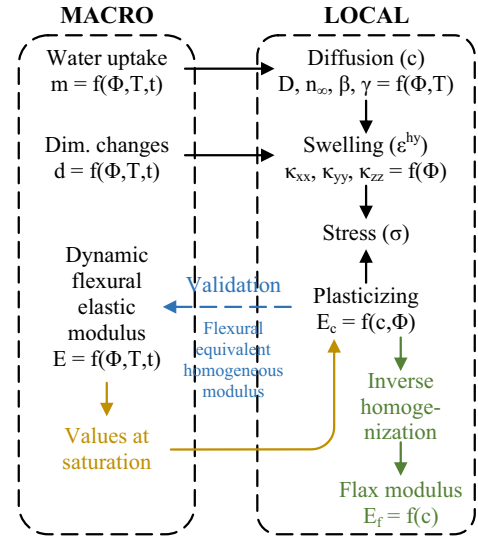

**Fig. 4.** Schematic representation of the construction of the model proposed in this study.

– one of them displayed an equilibrium corresponding to a saturation of the polymer (PLA at 20 and 35 °C). The sorption behavior of PLA at 20 and 35 °C being very similar, the two curves are mostly overlapping. This behavior is called Fick diffusion [14],
– another one displayed a decrease of the sorption rate without reaching an equilibrium (PLA and PLA-F10 at 50 °C, PLA-F10, and PLA-F30 at 20 and 35 °C). This behavior is called pseudo-Fick diffusion or two-stage diffusion [15],

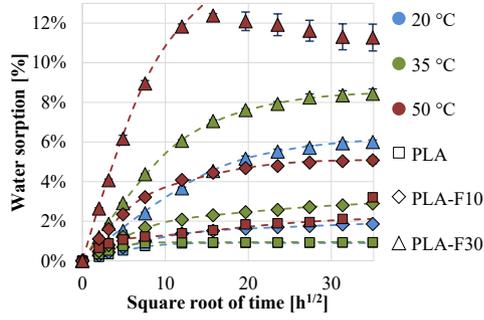

**Fig. 5.** Experimental (markers) and simulated (dashed lines) water content of PLA and flax/PLA composites during immersion in water at different aging temperatures.

- the latter displayed a maximum before a slow decrease of the water content (PLA-F30 at 50 °C). This behavior is typical of chemical damage.

These diffusion kinetics can be modeled by partial differential equations. However, in this first attempt, the models did not take into account any kind of damage. Consequently, only the first two kinetics could be correctly simulated.

The model used to represent Fick diffusion is the analogy of the heat flux equation [16]:

$$\frac{dc}{dt} = D \cdot \Delta c \quad (1)$$

where
$c$ is the local water content,
$D$ the diffusion coefficient (or diffusivity),
$\Delta$ the Laplacian operator.

While, the "two-stage diffusion" model implies the use of two variables for water content [17]:

$$\begin{cases} \frac{\partial n}{\partial t} = D \cdot \Delta n - \frac{\partial N}{\partial t} \\ \frac{\partial N}{\partial t} = \gamma n - \beta N \\ c = n + N \end{cases} \quad (2)$$

where
$n$ is the free water molecules content,
$N$ the bound water molecules content,
$\gamma$ the probability per unit time of free molecules to become bound,
$\beta$ the probability per unit time of bound molecules to become mobile.

As a result, Fick diffusion is a specific case of two-stage diffusion, where $N = 0$. So, the latter was chosen to simulate every experimental sorption results.

The overall water content in the geometry was obtained by integration:

$$\frac{M(t)}{M_0} = \frac{\int_V [n(t) + N(t)] dV}{\int_V dV} \quad (3)$$

### 3.2. Swelling

When it comes to swelling modeling, local strain was mostly considered to be proportional to local water content, as analogy to thermal expansion:

$$\boldsymbol{\varepsilon}^{hy} = \boldsymbol{\kappa} \cdot c \quad (4)$$

where

$\boldsymbol{\varepsilon}^{hy}$ is the hydric strain tensor,
$\boldsymbol{\kappa}$ the swelling tensor.

Swelling curves may display a gradual speed increase at early beginning, meaning locally a delay of swelling compared to sorption due to filling of free volume [15]. It is possible to introduce a swelling triggering water content from which the material starts to swell. In this case, the equation becomes:

$$\boldsymbol{\varepsilon}^{hy} = \boldsymbol{\kappa} \cdot (c - c_t) \quad \text{for } c \geqslant c_t \quad (5)$$

where $c_t$ is the swelling triggering water content (typically around 0.1%).

As expected, experimental results highlighted different swelling amounts according to each space direction (c.f. Section 4.2). Since swelling turned out to be anisotropic, an orthotropic swelling tensor was chosen: this assumption was dictated by the symmetry of the injected parallelepiped specimen

$$\boldsymbol{\varepsilon}^{hy} = \begin{bmatrix} \kappa_{xx} & 0 & 0 \\ 0 & \kappa_{yy} & 0 \\ 0 & 0 & \kappa_{zz} \end{bmatrix} \cdot (n + N - c_t) \quad (6)$$

This tensor was defined by three swelling coefficients $\kappa_{xx}, \kappa_{yy}, \kappa_{zz}$ which characterize the swelling in the x, y, and z direction respectively.

### 3.3. Elasticity

For the mechanical behavior, this study focused on the modeling of the elastic behavior of materials during diffusion (i.e. as a function of the water content at the local scale). Consequently, the Hooke's law was used for the linear behavior:

$$\boldsymbol{\sigma} = [C] : \boldsymbol{\varepsilon}^{el} \quad (7)$$

where
$\boldsymbol{\sigma}$ is the stress tensor,
$[C]$ is the stiffness matrix,
$\boldsymbol{\varepsilon}^{el}$ is the elastic strain tensor.

Literature revealed complex fiber orientation mechanisms during injection molding. The authors refer in this respect the reader to the source document relative to the induced anisotropy distribution of the fibers in the injected structures [18]. However, without prior knowledge of the quantitative distribution of fibers and in order to propose a model easy to implement and independent of the geometry, it was decided to adopt an isotropic elastic behavior. As a result, the expression of the stiffness matrix is as follow:

$$[C] = \frac{E}{(1+\nu)(1-2\nu)} \begin{bmatrix} 1-\nu & \nu & \nu & & & \\ \nu & 1-\nu & \nu & & 0 & \\ \nu & \nu & 1-\nu & & & \\ & & & \frac{1-2\nu}{2} & 0 & 0 \\ & 0 & & 0 & \frac{1-2\nu}{2} & 0 \\ & & & 0 & 0 & \frac{1-2\nu}{2} \end{bmatrix} \quad (8)$$

The stiffness matrix is proportional to the Young's modulus (E) which depended on the water content. Besides, the Poisson's ratio ($\nu$) was supposed to be independent to the water content.

The static fundamental principle states stress equilibrium:

$$\text{div} \boldsymbol{\sigma} = 0 \quad (9)$$

In order to take into account swelling, the total strain tensor ($\boldsymbol{\varepsilon}^t$) was introduced:

$$\boldsymbol{\varepsilon}^t = \boldsymbol{\varepsilon}^{el} + \boldsymbol{\varepsilon}^{hy} \quad (10)$$

This equation enabled to obtain a generalized Hooke's law:

$$\boldsymbol{\sigma} = [C] : (\boldsymbol{\varepsilon^t} - \boldsymbol{\varepsilon^{hy}}) \quad (11)$$

### 3.4. Plasticizing

To complete the model, the dependence of the local elastic modulus of materials to the water content had to be determined. It was made possible by assessing the macroscopic modulus of samples for different homogeneous water contents achieved by exposing samples to different environmental conditions (c.f. Table 1). For each material, it enabled the association of a local elastic modulus to its water content (markers in Fig. 3). These experimental moduli were interpolated by a third degree polynomial function for each material. These functions were implemented in the numerical model to provide local moduli depending on the local water content.

A macroscopic simulated modulus was defined from the local modulus distribution. In order to enable the comparison with the experimental dynamic elastic modulus, it was evaluated in flexion. This flexural equivalent homogeneous modulus ($E_{eq}$) was obtained from the integration of the local elastic modulus over the cross-section located at the mid-length of the sample. The expression of $E_{eq}$ is:

$$E_{eq} = \frac{\int_S z^2 E(y,z) dS}{\int_S z^2 dS} \quad (12)$$

where
  $S$ is the section (in y:z plane) located in the middle of the sample,
  $E$ is the local Young's modulus.

### 3.5. Implementation in the numerical model

The geometry used for the finite element model represented one eighth of the sample according to the three planes of symmetry (c.f. Fig. 1).

Meshing of the geometry was obtained firstly by a regular mesh (53 × 7 elements) of the upper side with quadrilateral plane elements. Then, these elements were extruded along thickness axis leading to four layers of hexahedral brick element. Finally, it turned out to be necessary to refine elements near the sides exposed to environmental conditions.

These peripheral sides were assigned to a constant free water molecule content of $n_\infty$ and boundary conditions of free displacement were defined. The other sides, corresponding to the planes of symmetry, were assigned with no flux of water and no displacement in the orthogonal direction.

As shown in Fig. 4, the model construction was successive. The first step consisted in modeling water diffusion. The identification of diffusion parameters (i.e. $c_t$, D, β, γ, and $n_\infty$ the free water molecule content at saturation) for each material and each temperature was tuned by minimizing the difference between experimental and simulated water uptakes. As swelling is a direct consequence of the presence of water, this phenomenon was the second to be modeled. Based on the water content distribution, the swelling parameters (i.e. $\kappa_{xx}$, $\kappa_{yy}$, $\kappa_{zz}$) were tuned using the same approach with dimensional measurements assessed in the three directions. Finally, the water-dependent elastic behavior was added to the model, on the basis of the moduli measured at the different saturation levels (c.f. Section 2.2.3). A flexural equivalent homogeneous modulus was defined taking into account the heterogeneous distribution of water. Its evolution during aging was compared to experimental measurements so as to verify the validity of the model.

### 3.6. Fiber stiffness dependency to water content

Based on these results, it was proposed to assess the stiffness of flax fibers as a function of their water content, thus enabling the simulation of the evolution of the elastic modulus of composites whatever their fiber content. This assessment was made possible by implementing the elastic behavior of PLA and composites at saturation as a function water content into a proper homogenization model. The assumption of an isotropic elastic behavior (cf. Section 3.3) implied a random distribution and orientation of fibers in the 3 space directions. As a result, the Halpin-Kardos model was used for homogenization [19]:

$$E_c = 0.184 E^L + 0.816 E^T \quad (13)$$

$$E^X = \frac{1 + \eta V_f \zeta^X}{1 - \eta V_f} E_m \quad (14)$$

$$\eta = \frac{E_f - E_m}{E_f + \zeta^X E_m} \quad (15)$$

where
  $E_m$ is the matrix elastic modulus,
  $E_f$ is the fiber elastic modulus,
  $V_f$ is the fiber volume fraction,
  $\zeta^X$ is a reinforcement geometry parameter; $\zeta^L = 2\phi$ and $\zeta^T = 2$,
  $\phi$ is the fibers shape factor (the relation between the length and the diameter of short fibers) determined in Section 2.2.2.

The modulus obtained by homogenization provides a local evaluation of the stiffness of a composite for a given fiber content. Finally, the water content dependency of the fiber modulus was obtained by matching the model to the measured composite moduli. In practice, the fit curve (dotted lines) of PLA in Fig. 3 was implemented in Eqs. (13)–(15) to plot the moduli of composites as a function of water content. By manually adjusting the dependency of the flax modulus as a function of water content, it was possible to fit the moduli of both composites as a function of water content to experimental data in Fig. 3.

## 4. Results and discussion

### 4.1. Water sorption/diffusion

The fitting of weight gain plots with the two-stage diffusion model is represented in Fig. 5. Results showed a very good correlation between the model and the experimental data, except for the PLA-F30 aged at 50 °C. Indeed, in this case, aging temperature is very near the glass transition temperature of the resin inducing unexpected phenomena such as the weight loss observed that has not been taken into account by this model.

The four parameters of the diffusion model (i.e. the diffusivity D, the free water content at saturation $n_\infty$, and the parameters β and γ) introduced in Section 3.1 and determined by adjustment are presented in Table 2. Diffusivity as well as saturation water content increased with temperature, and more drastically when it was close from glass transition temperature. This result can be qualitatively compared to many observations performed by other authors on epoxy systems [20,21]. Concerning the fiber content, its influence was different according to the considered parameter. At first, it might be surprising to notice that a higher content reduced the diffusivity. Yet it can be explained by the fact that it quantifies the rapidity of sorption to reach equilibrium. But saturation was achieved later with higher fiber content. This observation was in accordance with a previous study on PET/hemp composites which showed a diffusivity of fibers much lower than the matrix [22]. Besides, to our knowledge, the diffusivity of flax fibers in immersion remains unknown, it would have been difficult to obtain a homogenization of the composites diffusivity regarding the imperfect attempts present in literature [23,24].

**Table 2**
The four parameters of the two-stage diffusion model (D, $n_\infty$, β, and γ) for PLA and flax/PLA composites during immersion in water at different aging temperatures.

| Parameter | D [m²/s] | | | $n_\infty$ [%] | | | β [1/s] | | | γ [1/s] | | |
|---|---|---|---|---|---|---|---|---|---|---|---|---|
| Temperature | 20 °C | 35 °C | 50 °C | 20 °C | 35 °C | 50 °C | 20 °C | 35 °C | 50 °C | 20 °C | 35 °C | 50 °C |
| PLA | 2.30E−12 | 5.60E−12 | 1.20E−11 | 0.90% | 0.93% | 1.10% | 1.0e−8 | 9.0e−8 | 6.0e−7 | 2.0e−10 | 8.0e−9 | 6.0e−7 |
| PLA-F10 | 1.30E−12 | 2.20E−12 | 3.60E−12 | 1.51% | 2.05% | 3.40% | 2.0e−7 | 1.2e−6 | 1.2e−6 | 9.0e−8 | 2.3e−7 | 6.0e−7 |
| PLA-F30 | 6.20E−13 | 1.30E−12 | 2.70E−12 | 5.25% | 6.5% | 10.0% | 4.2e−7 | 9.0e−7 | 1.8e−6 | 9.0e−8 | 2.8e−7 | 9.0e−7 |

### 4.2. Swelling

Simulated swelling was evaluated in the (y:z) plane, for thickness and width, and in the (x:z) plane, for length (c.f. Fig. 1). In a given direction, swelling turned out to be heterogeneous due to the diffusion gradients. That is why the simulated swelling was calculated from the average strain of the finite elements located at the surface of samples where dimensional measurements were undertaken.

Swelling coefficients $\kappa_{xx}$, $\kappa_{yy}$, $\kappa_{zz}$ were determined by minimizing the difference between simulated and experimental swelling at 20 °C. Indeed, at this temperature, dimensional changes were shown to be completely reversible, and hence only induced by hygroscopic strains [11]. The resulting water swelling coefficients are represented in Fig. 6.

It is reported in literature that swelling of flax fibers is particularly anisotropic. While it is very low in their longitudinal direction, strain can reach 50% in their transverse direction [25]. As a result, the evolution of swelling coefficients with fiber content was strongly linked to the orientation of fibers in the matrix. That is why the orientation of fibers in the direction of the molding flow during injection molding process [18] was responsible for the increase (y and z directions) and the decrease (x direction) of swelling coefficients with fiber content.

Concerning the highest swelling coefficient in the z direction, it could have been caused by a possible highest skin/core ratio along the thickness than along the width. Indeed fibers were very likely oriented by the molding flow when they were close to mold walls. Based on the assumption that the afflicted area (skin) was relatively independent of the distance between two opposite walls, skin/core ratio should be higher for a narrow channel.

The influence of temperature on swelling was noteworthy. Tables 3 and 4 show experimental and simulated dimensional changes respectively in thickness and length. The behavior in the width direction is not presented in this study but is very similar to the thickness behavior.

At 20 °C, for each direction, simulated swelling was in accordance with experimental results, since water swelling coefficients were calibrated on those results. The temperature-dependent swelling behavior was different according to fiber content.

For PLA, when temperature was close to its glass transition temperature, simulated values were over-estimated for length and under-estimated for thickness. At 50 °C, this severe retraction in longitudinal direction, compensated by an increased swelling in transverse directions, was attributed to chains relaxation mechanisms caused by the drastic increase of chains mobility.

For composites, this irreversible behavior was reduced with the fiber content due the more stable structure conferred by the fibrous network. That is why in given conditions, both the standard deviations of experimental measurements and the differences between simulated values and experimental measurements decreased with fiber content.

### 4.3. Elasticity

The third degree polynomial interpolations of the local dynamic elastic modulus of PLA and flax/PLA composites as a function of the water content are presented in Fig. 3 along with experimental data (obtained both in immersion and in atmospheric conditions). Their implementation in the numerical model permitted the evaluation of the flexural equivalent homogeneous modulus as a function of aging time on the account of the distribution of water previously calculated.

These results are presented in Fig. 7 and compared with experimental dynamic elastic moduli. The progressive behavior of PLA was properly modeled for the three temperatures studied. Concerning the PLA-F10, the quality of the simulation depended on the considered temperature. Indeed, discrepancies only appeared at 35 °C for long term immersion, and at 50 °C for short term immersion. In both cases, the simulated moduli were overestimated. However, for PLA-F30, the simulated moduli were underestimated but only in the transient region. These inaccuracies of the simulated values were explained by the model approximations and more specifically by the absence of consideration of irreversible mechanisms [11]. Nevertheless, underestimated values were more likely due to an inaccurate evaluation of the water gradient or an inappropriate estimation of the flexural equivalent homogeneous modulus.

### 4.4. Fiber stiffness dependency to water content

In order to evaluate the influence of plasticizing on flax fibers, the dependence of their elastic modulus to water content was updated to fit the local Halpin-Kardos homogenized moduli of PLA-F10 and PLA-F30 to the experimental values determined on saturated samples. The dependence of the elastic modulus of flax fibers to water content was also modeled by a third degree polynomial function. The resulting behavior of flax fibers as a function of water content is presented in Fig. 8. Yet the dependency of modulus to water could be slightly overestimated. Firstly, the hypothesis of a homogeneous material would suggest that the actual water content in fibers was more significant based on the low sorption of PLA (cf. Fig. 5). Secondly, the occurrence of damage at the fiber/matrix interface [11] would result in a overestimation of the fibers modulus for high water contents. Nevertheless these results are globally in accordance to data found in the literature [26–28].

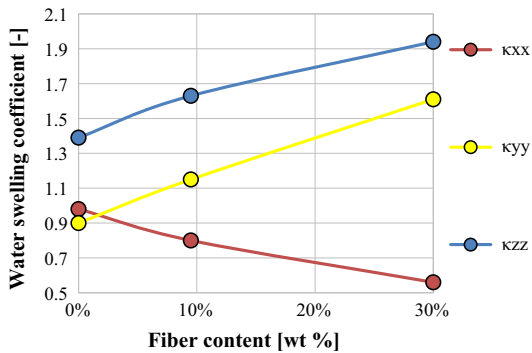

**Fig. 6.** Water swelling coefficients depending on each space direction for ISO 1BA geometry (c.f. Fig. 1) of PLA and flax/PLA composites.

**Table 3**
Experimental and simulated swelling in z direction (thickness) of PLA and flax/PLA composites after 144 h immersion in water at different aging temperatures.

| Aging temperature | 20 °C | | | 35 °C | | | 50 °C | | |
|---|---|---|---|---|---|---|---|---|---|
| Sample | PLA | PLA-F10 | PLA-F30 | PLA | PLA-F10 | PLA-F30 | PLA | PLA-F10 | PLA-F30 |
| Experimental swelling | 0.29% ± 0.04% | 0.76% ± 0.07% | 2.96% ± 0.09% | 0.36% ± 0.05% | 1.56% ± 0.10% | 4.99% ± 0.22% | 7.91% ± 0.54% | 5.59% ± 0.20% | 8.23% ± 0.18% |
| Simulated swelling | 0.31% | 0.77% | 2.96% | 0.33% | 1.18% | 4.10% | 0.74% | 2.07% | 7.28% |
| Difference | −0.02% | −0.01% | 0.00% | 0.03% | 0.38% | 0.89% | 7.17% | 3.52% | 0.95% |

**Table 4**
Experimental and simulated swelling in x direction (length) of PLA and flax/PLA composites after 144 h immersion in water at different aging temperatures.

| Aging temperature | 20 °C | | | 35 °C | | | 50 °C | | |
|---|---|---|---|---|---|---|---|---|---|
| Sample | PLA | PLA-F10 | PLA-F30 | PLA | PLA-F10 | PLA-F30 | PLA | PLA-F10 | PLA-F30 |
| Experimental swelling | 0.20% ± 0.03% | 0.40% ± 0.06% | 0.87% ± 0.12% | −0.02% ± 0.04% | 0.51% ± 0.04% | 0.86% ± 0.06% | −5.43% ± 0.76% | 0.47% ± 0.25% | 0.68% ± 0.05% |
| Simulated swelling | 0.22% | 0.38% | 0.85% | 0.23% | 0.58% | 1.18% | 0.52% | 1.02% | 2.10% |
| Difference | −0.02% | 0.02% | 0.02% | 0.25% | −0.07% | −0.32% | −5.95% | −0.55% | −1.42% |

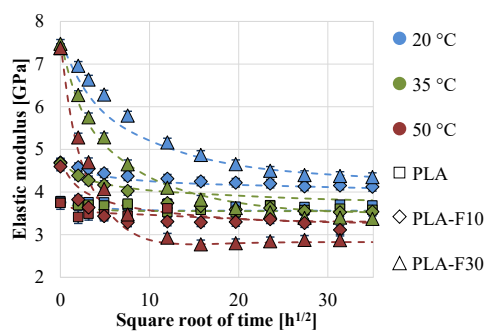

**Fig. 7.** Experimental dynamic elastic modulus (markers) and simulated flexural equivalent homogeneous modulus (dashed lines) of PLA and flax/PLA composites during immersion in water at different aging temperatures.

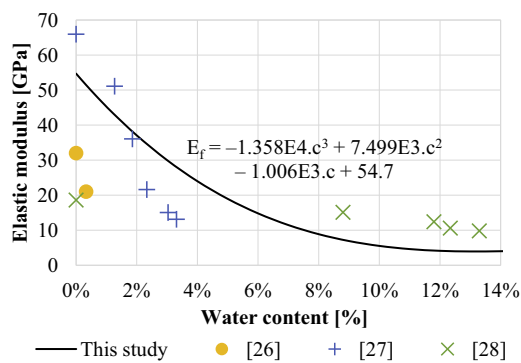

**Fig. 8.** Adjusted values of the local elastic modulus of flax fibers as a function of water content compared to literature data.

## 5. Conclusion

The purpose of this study was to simulate the reversible effects of the hydrothermal aging of flax reinforced composites. The heterogeneous mechanisms induced by hydrothermal aging, i.e. diffusion, swelling and plasticizing, were successively implemented in a numerical model which enabled taking into account the three dimensional geometry of samples. Parameters of the model were adjusted to match the experimental data.

Results showed that diffusion and swelling were perfectly simulated as long as the effects of aging were actually reversible, i.e. at 20 and 35 °C for all materials except PLA-F30 at 35 °C. In other conditions, the proximity of the glass transition of PLA combined with high water content triggered irreversible mechanisms such as polymer relaxation and hydrolysis. These mechanisms led to underestimated simulated swelling in samples transverse directions and overestimated simulated swelling in the samples longitudinal direction, thus highlighting the importance of the residual stress state resulting from processing. The determination of a flexural equivalent homogeneous modulus allowed verifying the dependence of the local dynamic elastic modulus of PLA and flax/PLA composites as a function of the water content. The evolution of this flexural equivalent homogeneous modulus as a function of aging time was globally in agreement with experimental data. Otherwise, the modulus decrease was generally overestimated for short-term (transient state) and slightly underestimated for long-term (almost stationary state).

Finally, the elastic modulus of flax fibers as a function of water content was estimated by updating the model as to match the local Halpin-Kardos homogenized modulus of composites as a function of water content to the experimental data. Results were in accordance to previous studies in literature.

## Appendix A. Supplementary material

Supplementary data associated with this article can be found, in the online version, at http://dx.doi.org/10.1016/j.compositesa.2016.08.011.